\documentclass[aps,prd,preprint,superscriptaddress,showpacs,floatfix,
preprintnumbers,nofootinbib, a4paper]{revtex4-1}
\usepackage{color}
\usepackage{graphicx}
\usepackage{textcomp}
\usepackage{amsmath,amssymb,array}
\usepackage{enumerate}
\newcommand{\mathsym}[1]{{}} 

\def\gsim{\:\raisebox{-1.1ex}{$\stackrel{\textstyle>}{\sim}$}\:}

\newcommand{\ba}{\begin{array}} 
\newcommand{\ea}{\end{array}}
\newcommand{\be}{\begin{equation}}
\newcommand{\ee}{\end{equation}}
\newcommand{\beqa}{\begin{eqnarray}} 
\newcommand{\eeqa}{\end{eqnarray}}
\def\321{$SU(3)\times SU(2)\times U(1)$}

\begin{document} 
\vspace*{1cm}
\title{Interpreting 750 GeV diphoton excess in SU(5) grand unified theory} 
\bigskip 
\author{Ketan M. Patel}
\email{ketan@iisermohali.ac.in} 
\affiliation{Indian Institute of Science Education and Research Mohali, Knowledge City, Sector  81, S A S Nagar, Manauli 140306, India}
\author{Pankaj Sharma}
\email{pankaj.sharma@adelaide.edu.au} 
\affiliation{Center of Excellence in Particle Physics (CoEPP), The University of Adelaide, South Australia \vspace*{1cm}}

\begin{abstract}
The ATLAS and CMS experiments at the LHC have found significant excess in the diphoton invariant mass distribution near 750 GeV. We interpret this excess in a predictive nonsupersymmetric SU(5) grand unified framework with a singlet scalar and light adjoint fermions. The $750$ GeV resonance is identified as a gauge singlet scalar. Both its production and decays are induced by 24 dimensional adjoint fermions predicted within SU(5). The adjoint fermions are assumed to be odd under $Z_2$ symmetry which forbids their direct coupling to the standard model fermions. We show that the observed diphoton excess can be explained with sub-TeV adjoint fermions and with perturbative Yukawa coupling. A narrow width scenario is more preferred while a simultaneous explanation of observed cross section and large total decay width requires some of the adjoint fermions lighter than 375 GeV. The model also provides a singlet fermion as a candidate of cold dark matter. The gauge coupling unification is achieved in the framework by introducing color sextet scalars while being consistent with the proton decay constraint.
\end{abstract} 

\preprint{ADP-16-15/T970}
\maketitle

\section{Introduction} 
\label{intro}
The first set of data collected by ATLAS and CMS experiments at the Large Hadron Collider (LHC) with proton-proton collisions at center-of-mass energy $\sqrt{s} = 13$ TeV has recently been presented by the respective collaborations \cite{exp}. Both the experiments claim an excess in the diphoton channel at the invariant mass near 750 GeV. This result is based on 3.2 fb$^{-1}$ and 2.6 fb$^{-1}$ of integrated luminosities collected by the ATLAS and CMS respectively. It has the local statistical significance of 3.9$\sigma$ and 2.6$\sigma$ respectively. The excess around the same mass has not been seen in any other channels like dijet, dilepton or diboson. Also, the similar excess was not seen even in diphoton channel in the previous run with $\sqrt{s} = 8$ TeV.  The new excess is best described by relative width $\Gamma/M \approx 0.06$ as suggested by the ATLAS collaboration while the same hypothesis by the CMS collaboration reduces the significance of excess to 2.0$\sigma$ \cite{exp}. Based on the claimed width, the estimated cross section is given as \cite{exp, Franceschini:2015kwy}
\beqa \label{csgamma}
\sigma(pp \to X \to \gamma \gamma) &\approx & (10 \pm 3)~ {\rm fb}~~~{\text{by ATLAS \cite{exp}}}~, \nonumber \\
 &\approx & (6 \pm 3)~ {\rm fb}~~~{\text{by CMS \cite{exp}}}~,\eeqa
where $X$ represents a new intermediate massive state with mass $M\approx 750$ GeV.

It is quite possible that the above experimental result is just a statistical fluctuation and more data may eventually wipe out this primitive signal. However, if confirmed, this will be the first direct observation of sub TeV scale physics beyond the Standard Model (SM). While the more LHC data will decide the fate of this signal, it is interesting to interpret this signal in the context of currently known new physics scenarios and if the viability of new models can be tested using this new resonance. Several proposals have already been forwarded in this direction \cite{Franceschini:2015kwy,Gupta:2015zzs,singletscalars,scalars,VLfermions,others}. A new particle decaying into two photons can be of either spin-0 or spin-2. Assuming $X$ as spin-0 $s$-channel resonance, as either a singlet or a weak doublet under the SM, the observed large cross section cannot be explained with the only SM degrees of freedom \cite{Gupta:2015zzs,singletscalars}. This includes class of models with only extra uncolored scalars including two Higgs doublet models. One of the modification to these simple scenarios is to add extra vector-like fermions which can contribute in both production and decay of this heavy scalar state \cite{Franceschini:2015kwy,Gupta:2015zzs,singletscalars,VLfermions}. The colored fermions with large electric charges can account for large cross section observed at the LHC. In typical bottom-up approaches suggested so far, a choice of such new vector-like fermions is quite adhoc and there exists many possible candidates to account for the diphoton signal \cite{singletscalars,VLfermions}.

In this paper, we present a model based on a well motivated class of nonsupersymmetric SU(5) grand unified theory \cite{Georgi:1974sy} and show that the diphoton excess can be accounted using  a scalar singlet, namely $S$ of mass $M=750$ GeV and light (sub TeV) fermions residing in a 24 dimensional adjoint representation of SU(5). The adjoint fermions arise in the SU(5) GUT with fixed gauge charges and one does not need to introduce adhoc vector-like fermions as it has been done in several works listed \cite{VLfermions}. We propose a renormalizable version of theory which provides predictive framework to account for diphoton excess in terms of finite number of vector-like fermions and with their unified Yukawa interactions with a scalar $S$. The adjoint fermions are kept odd under $Z_2$ symmetry and they have only gauge interactions and Yukawa interactions with $S$. All the tree level decays of $S$ into the SM fermions are forbidden by the gauge and discrete $Z_2$ symmetries. The scalar $S$ can be produced through gluon-gluon fusion at 1-loop and also can decay into two photons through similar triangle diagrams with adjoint fermions in the loop. We show that such a framework can account for large enough cross section of diphoton events observed at the LHC. We discuss both the narrow and broad width scenarios of diphoton excess provided by the different ranges of the masses of adjoint fermions. Because of the presence of an unbroken $Z_2$ symmetry, model also provides a candidate for fermionic cold dark matter. Further, we show how gauge coupling unification can be achieved in this model using light adjoint fermions and a couple of light colored scalar fields being consistent with the proton decay constraints. 

We describe the model in the next section. In section \ref{diphoton}, we discuss how the diphoton excess can be fitted in the model. The gauge coupling unification and constraints from proton decay are discussed in section \ref{unification}. Finally, we summarize in section \ref{summary}.

\section{The Model}
\label{model}
We assume that the 750 GeV resonance seen at the LHC is a singlet scalar under the SM. Implementing it in SU(5), we assume that it is also a singlet under SU(5). Its production and decays are mediated by the vector-like fermions which belong to an adjoint of SU(5), namely $24_F$. We impose a discrete $Z_2$ symmetry under which $24_F$ is odd while all the other fields are even. The SM fermions are accommodated in a fundamental and in an antisymmetric representations of SU(5) as $\overline{5}_F$ and $10_F$ respectively \cite{Georgi:1974sy}. The consistent gauge symmetry breaking down to the SU(3)$_{\rm C}\times$U(1)$_{\rm em}$ and viable charged fermion masses are generated by introducing $5_H$, $24_H$ and $45_H$ scalars. The $45_H$ is needed to remove the degeneracy between the masses of charged leptons and down-type quarks at the renormalizable level \cite{Georgi:1979df,Perez:2007rm}. The gauge and Yukawa interactions follow in the standard way \cite{Georgi:1974sy,Perez:2007rm}. The renormalizable scalar Lagrangian of the model can be written as 
\be \label{scalar}
{\cal L}_{\rm scalar}= \frac{M_S}{2} S^2 + \lambda S^4  + {\cal L}(5_H, 24_H, 45_H,S).
\ee
where ${\cal L}(5_H, 24_H, 45_H,S)$ is the Lagrangian including the scalar fields $5_H$, $24_H$ and $45_H$ written in standard way \cite{Georgi:1974sy,Perez:2007rm} with their additional interactions with singlet. The GUT symmetry breaking is induced by the vacuum expectation value (vev) of $24_H$ preserving the SM gauge symmetry. For this to happen, a suitable choice for vev is $\langle 24_H \rangle = {\rm Diag.} (2,2,2,-3,-3)v /\sqrt{30}$. The $5_H$ and $45_H$ each contains SM like Higgs doublets which get mixed with each other and one of the linear combinations remains light that can be identified as the SM Higgs boson. Note that a fine tuning is needed to arrange such a light Higgs doublet while keeping the other multiplets in $5_H$ and $45_H$ as heavy as the GUT scale.  We assume such fine tuning in the model. The minimal version of renormalizable SU(5) does not account for neutrino masses. This can easily be solved by adding either singlet fermion or $15_H$ \cite{Dorsner:2005fq} and inducing small neutrino masses through type I or type II seesaw mechanisms respectively.

The $24_F$ fermions have the standard gauge interactions. In the absence of $Z_2$ symmetry, they also have Yukawa interactions with the SM fermions through a gauge invariant term like $\overline{5}_F 24_F 5_H$. Such interactions result into the mixing between the SM fermions and adjoint fermions after electroweak symmetry breaking (EWSB) and lead to tree level decays of $S$ into the SM fermions. Since there already exists strong bounds on such decays from 8 TeV LHC \cite{Franceschini:2015kwy}, we forbid such interactions through a $Z_2$ symmetry under which $24_F$ is odd. Besides the usual kinetic terms, the renormalizable interaction of $24_F$ with the other fields in the model can be written as
\be \label{L24}
{\cal L}_{24_F}= m_F ~Tr(24_F^2) + \lambda_H ~Tr(24_F^2 24_H) + \lambda_S ~S~ Tr(24_F^2)~.
\ee
After SU(5) breaking through the vev of $24_H$, the first and second terms determine the mass spectrum of the various adjoint fermions residing in $24_F$ while the last term gives required interaction between adjoint fermions and a scalar of $M_S=750$ GeV. Under the SM gauge group, the $24_F$ decomposes as:
\be \label{24-decomp}
24_F = Q_8 (8,1,0) + Q_3 (3,2,-5/6) + \overline{Q}_3 (\bar{3},2,5/6) + L_3 (1,3,0) +L_1 (1,1,0). \ee
Here the first (second) index in the bracket indicates SU(3)$_{\rm C}$ (SU(2)$_{\rm L}$) representation of corresponding fermion while the last index is the hypercharge $Y$. The hypercharge is normalized such that the electric charge is given as $Q=T_{3\rm{L}}+Y$. Once the electroweak symmetry is broken, we have adjoint fermions $Q_8^0$, $L_1^0$, $L_3^0$, $L_3^\pm$, $Q_3^{\pm 1/3}$ and $Q_3^{\pm 4/3}$ where the superscript indicates the electric charge. All these fermions couple to the scalar $S$ and can contribute into the production and decay of $S$ through the 1-loop triangle diagrams.

Let us now discuss the mass spectrum of the $24_F$ fermions. All the multiplets within $24_F$ has a common mass $m_F$ which gets corrected after SU(5) is broken into the SM through a vev, $\langle 24_H \rangle = {\rm Diag.} (2,2,2,-3,-3) v /\sqrt{30}$. After EWSB, one gets
\beqa \label{24masses}
m_{Q_8} &=& m_F + \frac{2}{\sqrt{30}}\lambda_H v~, \nonumber \\
m_{Q_3,\overline{Q}_3} &=& m_F - \frac{1}{2\sqrt{30}}\lambda_H v~, \nonumber \\
m_{L_3} &=& m_F - \frac{3}{\sqrt{30}}\lambda_H v~, \nonumber \\
m_{L_1} &=& m_F - \frac{1}{\sqrt{30}}\lambda_H v~.\eeqa
Clearly, one can obtain the desired masses for any two multiplets using the free parameters while the masses of remaining fermions get fixed. All the multiplets however couple to $S$ with a universal coupling $\lambda_S$. Our aim in this paper is to investigate the viability of the fermions within $24_F$ in explaining the observed $\sigma(pp \to S \to \gamma \gamma)$. Before we carry out such an analysis in the next section, let us discuss below some salient features of the presented model.

\begin{itemize}
\item In order to explain the diphoton excess, the masses of adjoint fermions has to be much smaller than the GUT scale. One therefore has to assume that $M_S,~m_F \ll M_{\rm GUT}$ and $\lambda_H$ is vanishingly small for sub TeV $S$ and $24_F$.
\item The new 750 GeV scalar $S$ can be produced at the LHC dominantly through gluon fusion mediated by $Q_8^0$, $Q_3^{\pm 1/3}$ and $Q_3^{\pm 4/3}$. Its decay into a pair of photons is dominantly mediated by $Q_3^{\pm 4/3}$ and $L_3^{\pm 1}$. Hence the same set of adjoint fermions give rise to the observed $\sigma(pp \to S \to \gamma \gamma)$. This provides a very predictive framework for 750 GeV resonance since there is a unique coupling between various adjoint fermions and $S$ and the masses  of various adjoint fermions are also correlated as can be seen from Eq. (\ref{24masses}).
\item The neutral fermion $L_1^0$ can serve as a cold dark matter candidate in this model since its stability is guaranteed by an unbroken $Z_2$ symmetry. If the mass of $L_1^0$ is smaller than $M_S/2$, then $S$ can decay into a pair of $L_1^0$ giving missing transverse momentum signal at the LHC. As we show later, both the scenarios are open in which $L_1^0$ could be lighter or heavier than $M_S/2$. In the later case, the dark mattter can co-annihilate into the SM particles through $S$ and adjoint fermions. We leave further studies of dark matter abundance and constraints from direct and indirect search experiments for future investigations.
\item Note that $S$ can decay into a pair of the SM Higgs bosons through $\mu_H S 5_H^\dagger 5_H$ term in the scalar potential. In order to evade the current limits on such decays, one can assume small enough coupling $\mu_H$ without affecting the other phenomenological aspects of the model. The most stringent bound on this coupling currently comes from the diHiggs production at 8 TeV LHC \cite{bound:Shh}. In this model, the diHiggs production is dependent on two parameters, the Yukawa coupling $\lambda_S$ in the production of $S$, and coupling $\mu_H$ in the decay. For $\lambda_S=1.0$ and 0.5, the most stringent upper bounds on $\mu_H$ are estimated to be 100 GeV and 400 GeV respectively.
\item The $Z_2$ parity forbids the direct interactions between the $24_F$ and SM fermions. In the absence of such symmetry $24_F$ and $\overline{5}_F$ can have Yukawa interactions through $5_H$. This can lead to neutrino masses through type III (fermion triplet mediated) seesaw mechanism as discussed in \cite{Bajc:2006ia}. This scenario can also be a candidate model for diphoton excess however the constraints on $S$ decaying into the pair of SM fermions should be taken into account since such decays are now induced due to mixing between the $24_F$ and SM fermions.
\end{itemize}

\section{Fitting diphoton excess}
\label{diphoton}
The observed cross section for the diphoton excess is quite large $\sim 10$ fb. It is quite clear that this excess cannot be explained only through the contributions of the top quarks and $W$ bosons in the loops even if a new scalar is charged under the SM. Thus, this  resonance cannot be due to the only new scalar and should be accompanied by other new states such as the vector-like fermions \cite{VLfermions}. A model proposed in the previous section naturally contains such vector-like fermions that includes three colored particles, namely $Q_8^0$, $Q_3^{\pm 1/3}$ and $Q_3^{\pm 4/3}$ and one colorless charged particle $L_3^\pm$. Since $S$ is a singlet scalar, it does not have tree level interactions with the SM fermions and gauge bosons. Thus, it can only be produced through the gluon-fusion at loop level. Since it is a gluon-initiated loop, only colored particles will contribute in the production of $S$.

The partonic $gg\to S$ production can be written in the standard form in terms of $\Gamma(S\to gg)$ decay width as \cite{Djouadi:2005gi}:
\begin{equation}
 \hat{\sigma}(gg\to S)=\frac{\pi^2}{8M_S}\Gamma(S\to gg)\delta(\hat{s}-M_S^2),
\end{equation}
where, the decay width of $\Gamma(S\to gg)$ is given by \cite{Djouadi:2005gi,Franceschini:2015kwy}
\begin{equation}
 \Gamma (S\to gg) = M_S \frac{\alpha_s^2}{2\pi^3}\left|\sum_f C_{rf}\sqrt{x_f}\lambda_S \mathcal A(x_f)\right|^2.
\end{equation}
Here $C_3 = 1/2$ for the triplet and $C_8=3$ for the octet fermions. The $x_f=4M_f^2/M_S^2$, $\mathcal A(x_f)=1+(1-x_f)\arctan^2(1/\sqrt{x_f-1})$, and $M_f$ is the mass of adjoint fermion propagating in the loop.  The sum runs over all the colored states present in the model. The partonic cross section can easily be transformed into hadronic cross section by multiplying the integrated gluon-luminosities at 13 TeV. We find that the largest contribution to the production of $S$ comes from color octet $Q_8^0$ because of large factor of $C_8=3$ as compared to $Q_3^{\pm 1/3}$ and $Q_3^{\pm 4/3}$. Fermions $Q_3^{\pm 1/3}$ and $Q_3^{\pm 4/3}$, being color triplets, contribute equally to the production.
\begin{figure}[h!]
 \includegraphics[scale=0.85]{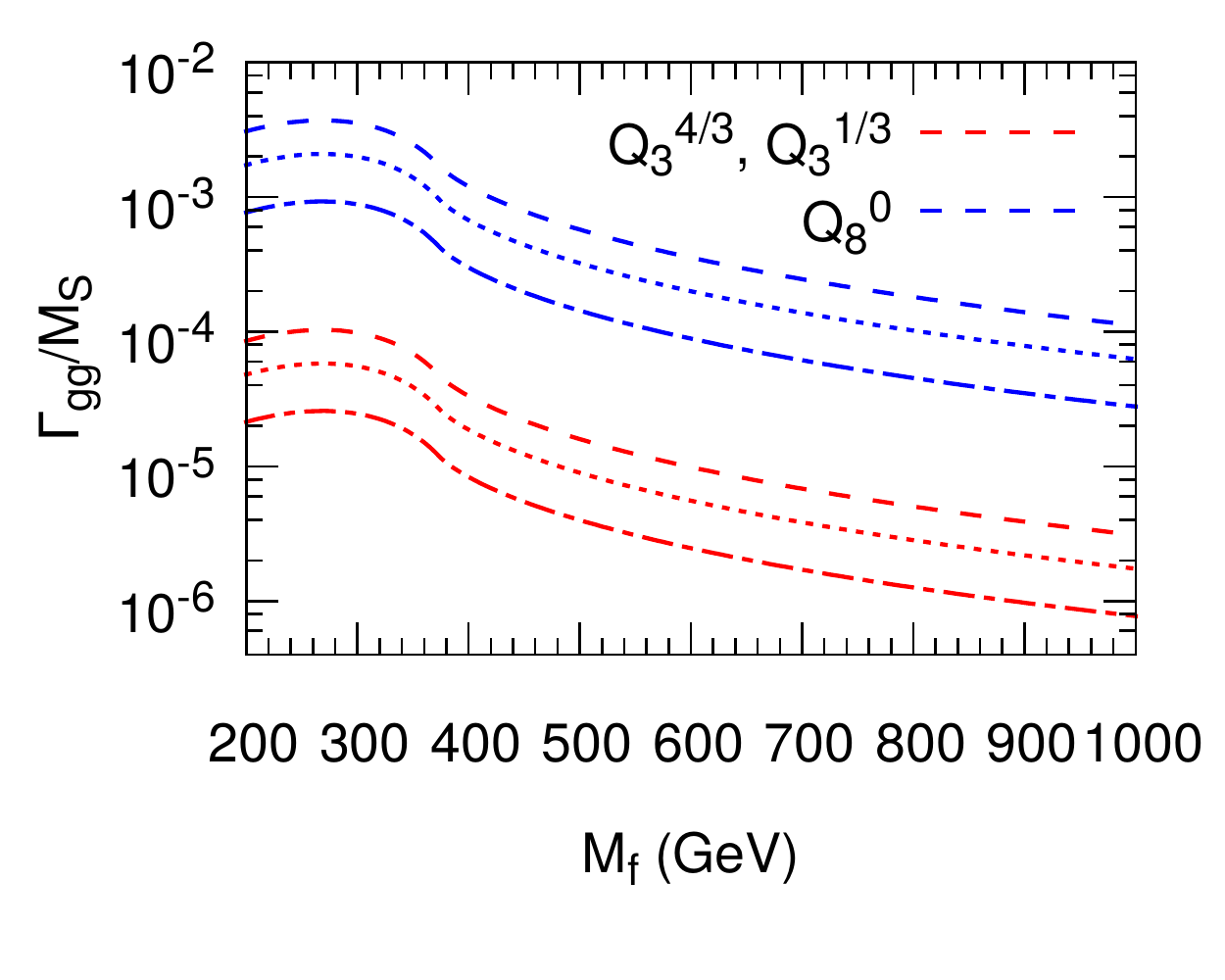}
 \caption{\label{fig:Gam_gg} The $\Gamma_{gg}/M_S$ for colored states: octet $Q_8$ (blue) and triplet $Q_3$ (red). Also shown are the contributions 
 for three values of Yukawa coupling $|\lambda_S|=$ 1.0 (long-dashed), 0.75 (short-dashed) and 0.5 (long-short dashed).}
\end{figure}
In Fig. \ref{fig:Gam_gg}, we show the contributions of $Q_8^0,~Q_3^{\pm 1/3}$ and $Q_3^{\pm 4/3}$ to $\Gamma_{gg}/M_S$ for three different values of Yukawa coupling $|\lambda_S|=$ 1.0 (long-dashed), 0.75 (small-dashed) and 0.5 (long-small dashed) respectively. The $\Gamma_{gg}/M_S$ is about an order magnitude larger for $Q_8^0$ than it is for $Q_3^{\pm 1/3}$ and $Q_3^{\pm 4/3}$.

The decay width of $S\to \gamma\gamma$ can be written as \cite{Djouadi:2005gi,Franceschini:2015kwy}
\begin{equation}
\Gamma (S\to \gamma\gamma) = M_S \frac{\alpha_{em}^2}{16\pi^3}\left|\sum_f d_{rf}Q_f^2\sqrt{x_f}\lambda_S \mathcal A(x_f)\right|^2,
\end{equation}
where for triplet $d_r=3$ and $Q_f$ is the electric charge of the particle. In the adjoint of SU(5), there are three electrically-charged particles. The largest contribution to diphoton decay width  will come from $Q_3^{\pm 4/3}$ followed by $L_3^\pm$ and $Q_3^{\pm 1/3}$. The contribution of $Q_3^{\pm 1/3}$ is nearly 1/9 of $L_3^\pm$ to diphoton width while that of $Q_3^{\pm 4/3}$ is 256/9 of $L_3^\pm$.
\begin{figure}[h!]
 \includegraphics[scale=0.85]{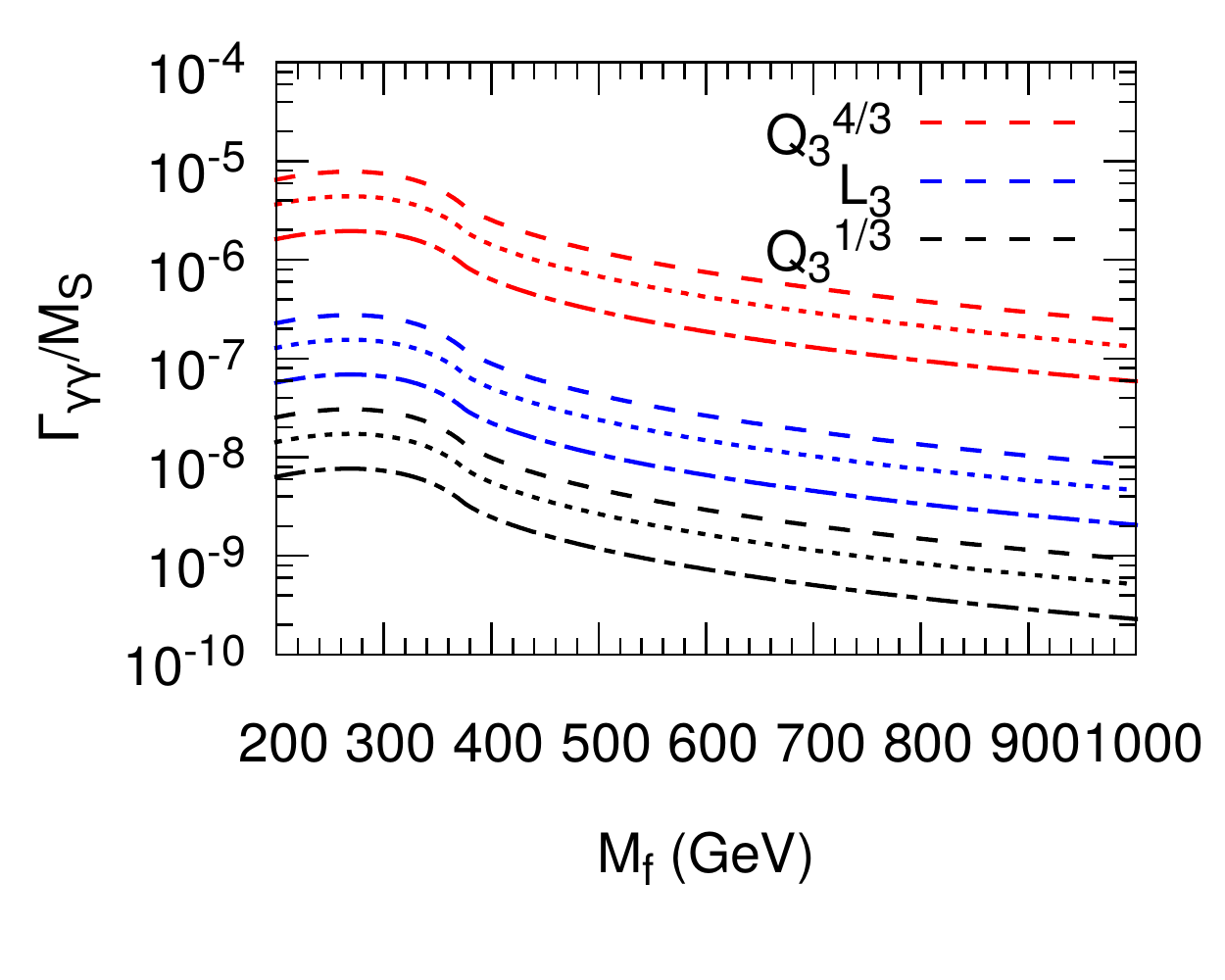}
 \caption{\label{fig:Gam_phph} The $\Gamma_{\gamma\gamma}/M_S$ for electrically-charged states: $Q_3^{\pm 4/3}$ (red), $Q_3^{\pm 1/3}$ (black) and $L_3^\pm$ (blue). 
Also shown are the contributions for three values of Yukawa coupling $|\lambda_S|=$ 1.0 (long dashed), 0.75 (short-dashed) and 0.5 (long-small dashed).}
\end{figure}
In Fig. \ref{fig:Gam_phph}, we show the contributions of $Q_3^{\pm 1/3}$, $Q_3^{\pm 4/3}$ and $L_3^\pm$ to $\Gamma_{\gamma \gamma}/M_S$ for three different values of the Yukawa coupling
$|\lambda_S|=$ 1.0 (long dashed), 0.75 (small-dashed) and 0.5 (long-small dashed).

\begin{figure}[h!]
 \includegraphics[scale=0.85]{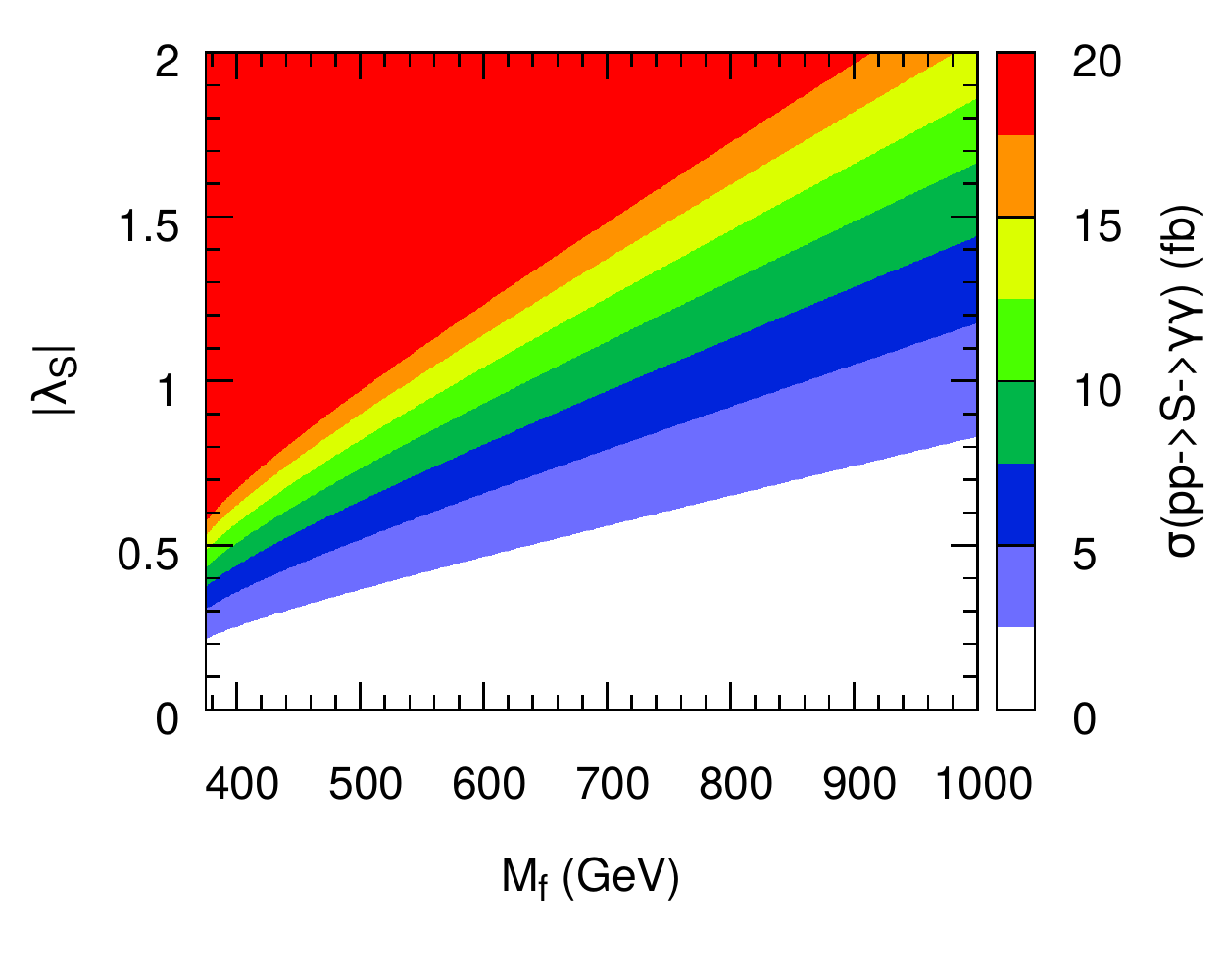}
 \caption{\label{fig:sig_gg} The production cross section of $S$ times the BR($S\to{\gamma \gamma}$) in the plane of common adjoint fermion mass $M_f$ and universal Yukawa coupling $|\lambda_S|$ for $M_f>M_S/2$.}
\end{figure}

We now discuss the cross section $\sigma(pp \to S \to \gamma \gamma)$ predicted in the model. In Fig. \ref{fig:sig_gg}, we show the production cross section of $S$ times the branching fraction in femtobarns in the plane of adjoint fermion masses and universal Yukawa coupling $|\lambda_S|$. For simplicity, we consider all adjoint fermions to be degenerate in masses and combine their contributions. The light blue region corresponds to 2.5-5 fb, the dark blue to 5-7.5 fb, dark green to 7.5-10 fb and light green to 10-12.5 fb of cross section of $pp\to S\to \gamma\gamma$. All these regions are compatible with the cross section measured by the ATLAS and CMS for the 750 GeV diphoton resonance as listed in Eq. (\ref{csgamma}). We also have implemented the model in {\tt MadGraph} \cite{mg5} using {\tt Feynrules} \cite{feynrules} and all the numbers for cross section and decay widths have been cross checked with {\tt MadGraph}.

The broad width scenario preferred by the ATLAS analysis requires the tree level decays of $S$. If the masses of adjoint fermions $M_f>M_S/2$, then $S$ decays only into $gg$ and $\gamma\gamma$ through loops. In such a scenario, the total width of the resonance is very small $\sim{\cal O}$(MeV). Note that $S\to hh$ decay is still open but it is induced by a coupling different than $\lambda_S$ as discussed in the previous section. We assume that such coupling is small enough and, thus, this channel is sub-dominant compared to $gg$ and $\gamma\gamma$ respecting the 8 TeV constraints on $S \to hh$. In such a scenario, only a few MeV of width can be achieved. Thus this scenario leads us to the narrow width of $S$ which is not yet conclusively disfavored by the data. The BR($S\to\gamma\gamma$) is much larger for this mass range of adjoint fermions which allows values of coupling $|\lambda_S|$ to be as small as $\sim \mathcal O(0.3)$ to explain the observed diphoton excess as it can be seen from Fig. \ref{fig:sig_gg}.

On the other hand, when $M_f<M_S/2$, the tree-level decays of $S$ into a pair of adjoint fermions open up and thereby increasing significantly the total width of $S$. Thus, the BR($S\to \gamma\gamma$) gets reduced in this mass range. As we discuss in the next paragraph, there exists direct search constraints on the masses of vector-like fermions from the LHC Run-I.  It allows a possibility in which $M_{Q^{\pm 1/3}_3} <  M_S/2$ while the rest of the charged particle masses are restricted to be higher than $M_S/2$. In such a scenario, the scalar $S$ would decay into a pair of $Q_3^{\pm 1/3}$ at the tree level. Thus the width of the $S$ around 45 GeV could also be achieved. 
\begin{figure}[h!]
\includegraphics[scale=0.63]{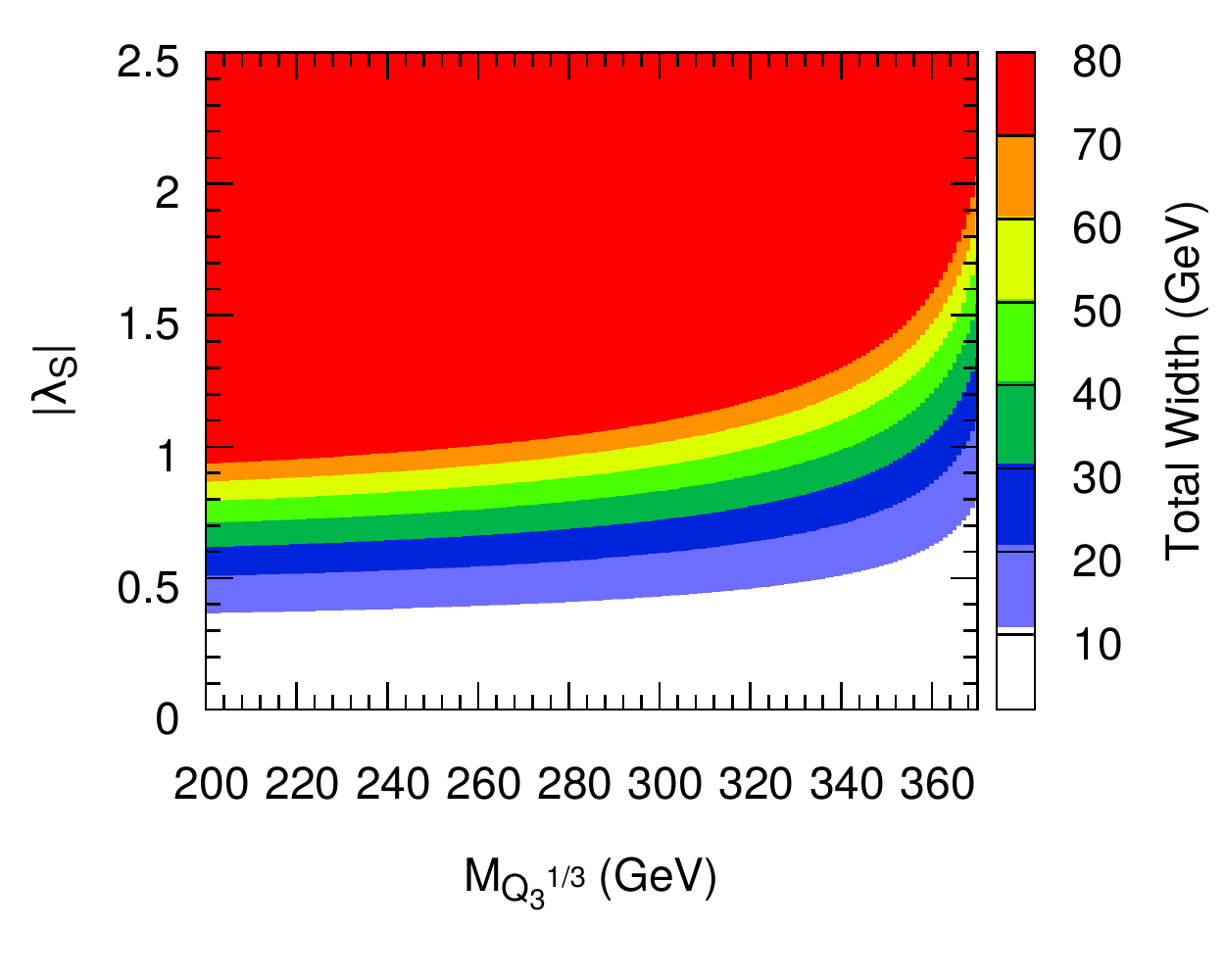}
\includegraphics[scale=0.63]{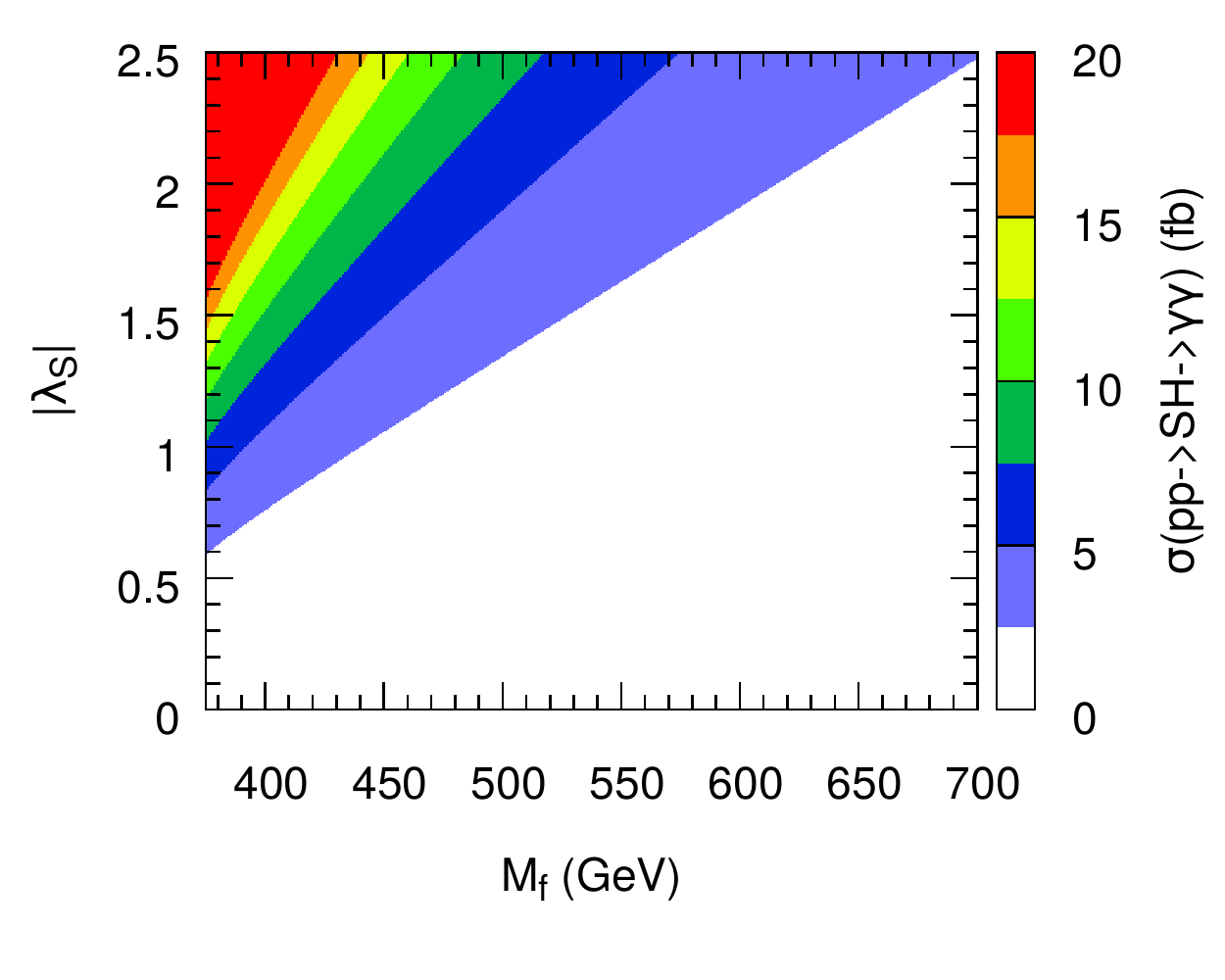}
\caption{\label{fig:width} Left Panel: The total decay width of the scalar $S$ in the plane of adjoint fermion mass $M_{Q_3^{\pm 1/3}}$ and universal Yukawa coupling $|\lambda_S|$ for $M_{Q_3^{1/3}}<M_S/2$. Right Panel: $\sigma(pp\to\gamma\gamma)$ in the plane of heavy fermion masses and Yukawa coupling $|\lambda_S|$. }
\end{figure}
In the left panel of Fig. \ref{fig:width}, we show the variation of total width of $S$ in the plane of $M_{Q_3^{\pm1/3}}$ and $|\lambda_S|$. We keep $M_{Q_3^{\pm 1/3}}<M_S/2$ and mass of all other fermions at 500 GeV. It can be seen that it is indeed possible to obtain the decay width upto 45 GeV for $0.8>|\lambda_S|>0.7$ and for $M_{Q_3^{\pm 1/3}}$ between 200 GeV-320 GeV. For the mass larger than 350 GeV, one needs relatively larger values of $|\lambda_s|$. In the right panel of Fig. \ref{fig:width}, we show the $\sigma(pp\to S \to \gamma\gamma)$ for this scenario in the plane of fermion mass $M_f$ and the Yukawa coupling $\lambda_S$. We keep $M_{Q_3^{1/3}} = 370$ GeV and a common mass $M_f$ for remaining fermions. In this case, because of the large tree-level decay width of $S\to Q_3^{1/3}Q_3^{-1/3}$, the BR($S\to\gamma\gamma$) gets reduced. Thus, one needs a large $\lambda_S$ coupling so that an appropriate diphoton cross section can be obtained. Clearly, the simultaneous explanation of large total decay width and large cross section requires very special values of masses for the vector-like fermions and universal Yukawa coupling. As can be seen from Fig. \ref{fig:width}, it can be achieved if  $M_{Q_3^{1/3}} \approx 370$ GeV, $M_f \approx 400$ GeV
and $|\lambda_S| \approx 1.2$.

The heavy adjoint fermions $Q_3$, $\overline{Q}_3$ and $L_3$ possess non-trivial SU(2)$_{\rm L}$ charges and hence they can induce decays like $S\to ZZ$, $S\to WW$ and $S \to Z\gamma$ at loop-level. We estimate the strengths of these decays in our model and compare them with the experimental constraints. In an effective theory at the electroweak scale, the leading order operators involving interactions between $S$ and SM gauge bosons can be written as
\be \label{gbeffective}
{\cal L} = \kappa_1 S B_{\mu\nu}B^{\mu\nu} + \kappa_2 S W^i_{\mu\nu}W^{i\mu\nu} + \kappa_3 S G^a_{\mu\nu}G^{a \mu\nu}~, \ee
where $G^a_{\mu \nu}$, $W^i_{\mu \nu}$ and $B_{\mu\nu}$ are field strength tensors of the gauge bosons of SU(3)$_{\rm C}$, SU(2)$_{\rm L}$ and U(1)$_{\rm Y}$ groups with $i=1,2,3$ and $a=1,..,8$. The couplings $\kappa_i = \kappa'_i {\mathcal A}(x_f)$ and $\kappa'_i$ for various adjoint fermions are listed in Table \ref{tab:eff_coup}. We use the low-energy Higgs theorem derived in \cite{Ellis:1975ap} to derive the coefficients $\kappa'_i$ for effective scalar-diboson vertex. We use  SU(5) normalization $g_Y = \sqrt{3/5}~ g_1$ where $g_1$ is the coupling which unifies with $g_2$ and $g_3$ at the unification scale.

\begin{table}[!ht]
\begin{center} 
 \begin{math} 
\newcolumntype{C}[1]{>{\centering\let\newline\\\arraybackslash\hspace{0pt}}m{#1}} 
\begin{tabular}{ |C{3.cm}||C{3.cm}C{3.cm} C{3.cm}   |}
\hline\hline 
~~~Fermions~~~  & ~~~~$\kappa'_1$~~~~ &~~~ $\kappa'_2$~~~ & ~~~$\kappa'_3$~~~ \\ \hline
$Q_8$ &  0 & 0 &$\frac{  \lambda_S}{8 \pi^2} \frac{g_3^2}{m_{Q_8}}$ \\
$(Q_3,~\overline{Q}_3)$ & $\frac{25 \lambda_S}{72 \pi^2} \frac{g_Y^2}{m_{Q_3}}$ & $\frac{ \lambda_S}{8 \pi^2} \frac{g_2^2}{m_{Q_3}}$ & $\frac{ \lambda_S}{12 \pi^2} \frac{g_3^2}{m_{Q_3}}$\\
$L_3$ & 0 & $\frac{ \lambda_S}{12 \pi^2} \frac{g_2^2}{m_{L_3}}$ & 0 \\
$L_1$ & 0 & 0 & 0\\
\hline
\hline 
\end{tabular}
 \end{math}
\end{center}
\caption{\label{tab:eff_coup} Effective couplings $\kappa'_i$ induced by various adjoint fermions in loop.} 
\end{table}
In the basis of physical gauge boson fields, the effective couplings between $S$ and various physical gauge bosons are given as
\beqa \label{effectivecouplings}
\kappa_{\gamma\gamma} &=& \kappa_1 \cos^2\theta_W+\kappa_2 \sin^2\theta_W~, \nonumber \\
\kappa_{ZZ} &=& \kappa_2 \cos^2\theta_W+\kappa_1 \sin^2\theta_W~, \nonumber \\
\kappa_{WW} &=& 2 \kappa_2~, \nonumber \\
\kappa_{Z\gamma} &=& (\kappa_2 - \kappa_1) \sin 2\theta_W~,\eeqa
where $\theta_W$ is the weak mixing angle. The ratio of the decay widths is simply determined as $\Gamma(S \to AB)/\Gamma(S \to \gamma \gamma) = |\kappa_{AB}|^2/|\kappa_{\gamma\gamma}|^2$. In the simple case of degenerate adjoint fermions {\it i.e.} $m_{Q_3}=m_{L_3}=M_f$, these ratios are estimated as 
\be \label{estimations}
\frac{\Gamma(S \to WW)}{\Gamma(S \to \gamma \gamma)} = 10.5,~~~~\frac{\Gamma(S \to ZZ)}{\Gamma(S \to \gamma \gamma)} = 2.0,~~{\rm and}~\frac{\Gamma(S \to Z\gamma)}{\Gamma(S \to \gamma \gamma)} = 0.5~.\ee
No resonance for 750 GeV scalar $S$ is seen in $WW$, $ZZ$ and $Z\gamma$ channels so far by the ATLAS and CMS. Using the $\sqrt{s}=8$ TeV ATLAS data and assuming the production cross section of $S$ grows as $r = \sigma_{13\rm{TeV}}/\sigma_{8\rm{TeV}} \approx 5$, the upper bounds on partial decay width of $S$ in various final states are derived at 95\% confidence level in Table I of \cite{Franceschini:2015kwy}. The current bounds are $\Gamma(S \to WW)/\Gamma(S \to \gamma \gamma) \le 20$, $\Gamma(S \to ZZ)/\Gamma(S \to \gamma \gamma)\le 6$ and $\Gamma(S \to Z\gamma)/\Gamma(S \to \gamma \gamma) \le 6$. The ratios estimated in Eq. (\ref{estimations}) assuming degenerate adjoint fermions are well within the current experimental bounds. Further, all the three ratios can be lowered if $m_L \gg m_{Q_3}$ without decreasing diphoton rate as the largest contribution in $\Gamma(S \to \gamma \gamma)$ arises from $Q_3$ and $\overline{Q}_3$.

We now briefly comment on the direct search constraints on vector-like fermions considered in this paper. The most stringent constraints on these fermions come from the searches of long-lived particles at the 7 and 8 TeV LHC by the CMS \cite{Chatrchyan:2013oca}. In these searches, the CMS collaboration  has utilized detector signatures like long time-of-flight to the outer muon system and anomalously high (or low) energy deposition in the inner tracker. These bounds are model independent and are only dependent on electric charges of the particles. Thus, constraints on the small electric charge particles are less stringent. As the electric charge increases, the bounds get stronger. For example, a lower bound on $Q=\pm 1/3$ particle is around 200 GeV, on $Q=\pm 1$ is around 400 GeV and on $Q=\pm 4/3$ is around 500 GeV. Therefore, only $Q_3^{\pm 1/3}$ can have mass smaller than $M_S/2$ among the adjoint fermions provided in the model. The bound on $Q_8^0$ can be obtained from long-lived gluino searches which is about 1.2 TeV. Considering these bounds, it is still possible to explain the observed cross section of diphoton events. As it can be seen from Fig. \ref{fig:sig_gg}, even for very heavy fermions of masses $\sim$ 1 TeV, the experimental result can be well explained albeit with relatively larger Yukawa couplings of $\sim \mathcal O(1)$. The narrow width for $S$ turns out to be more favorable in this case.

\section{Gauge coupling unification and proton decay}
\label{unification}
\begin{figure}[h!]
 \includegraphics[scale=0.6]{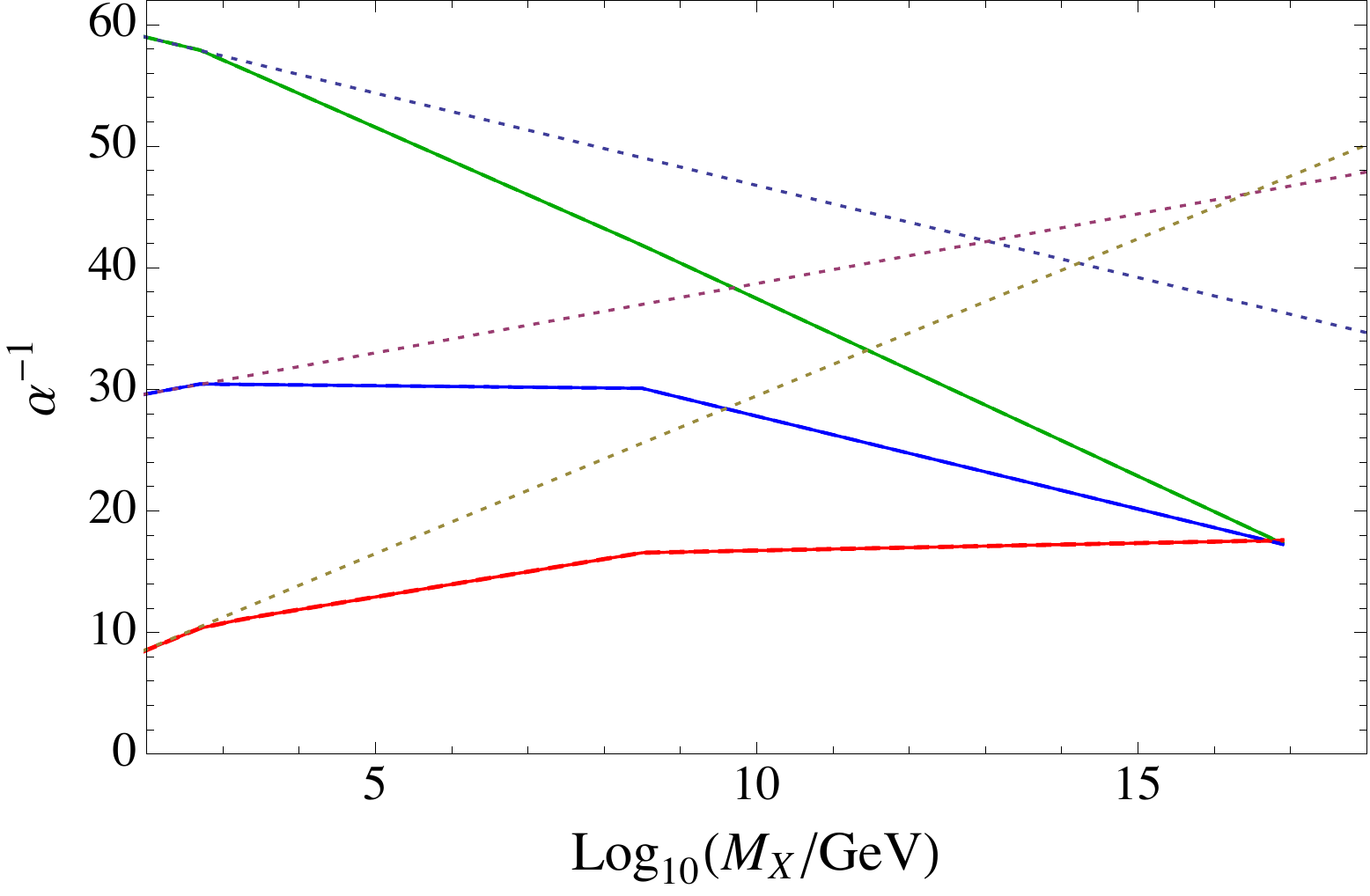}
 \caption{\label{fig:uni} Gauge coupling unification obtained by solving two loop RGE equations in the model (shown with solid lines) with degenerate adjoint fermions with mass 500 GeV and the scalars $(\overline{6},1,-1/3)$ and $(\overline{6},3,-1/3)$ with masses 2 TeV and $3.2 \times 10^{8}$ GeV respectively. The dotted lines correspond to running gauge couplings in the Standard Model.}
\end{figure}

Before discussing a possibility of gauge coupling unification in our framework, we briefly review the problem of unification in nonsupersymmetric GUT. As it is well known in the SM, the weak and strong gauge couplings unify at $\sim 10^{15}$ GeV which is an ideal scale considering the proton decay constraints. However, the U(1) gauge coupling meets the weak coupling at $\sim 10^{13}$ GeV spoiling the complete unification \cite{Langacker:1991an}. Adding a complete SU(5) multiplet below the GUT scale do not solve this problem as, at one loop of renormalization group evolution (RGE), it only changes the value of unified gauge coupling. We check that this problem continues at two loop also.  Hence the $24_F$ at the sub TeV scale do not really address the gauge coupling unification problem. One needs light incomplete set of multiplets to achieve gauge coupling unification. One such possibility was discussed by us in \cite{Patel:2011eh} where the unification was achieved with TeV scale color sextet scalars. We find that the same set of scalars can give gauge coupling unification together with light $24_F$. The new scalars transform as $(\overline{6},1,-1/3)$ and $(\overline{6},3,-1/3)$ under the SM gauge symmetry. The mass of weak singlet is required in the TeV range while the weak triplet can have mass in the range $10^{8}$-$10^{9}$ GeV \cite{Patel:2011eh}. For simplicity, we consider all the adjoint fermions residing in $24_F$ to be degenerate in masses and such mass can be anything in between $M_Z$ and $M_{\rm GUT}$, without spoiling the gauge coupling unification. As explained earlier, its main effect is in changing the value of the unified gauge coupling. Note that it is possible to obtain degenerate adjoint fermions assuming $\lambda_H$ vanishingly small in Eq. (\ref{24masses}). The existence of the gauge coupling unification is illustrated in Fig. \ref{fig:uni}.

We perform a two loop RGE analysis by choosing the scalars $(\overline{6},1,-1/3)$ and $(\overline{6},3,-1/3)$ with masses 2 TeV and $3.2 \times 10^{8}$ GeV respectively while the entire $24_F$ stays at 500 GeV. We have taken $\alpha_1(M_Z) = 0.016946 \pm 0.000006$,  $\alpha_2(M_Z) = 0.033812 \pm 0.000021$ and $\alpha_3(M_Z) = 0.1184 \pm 0.0007$ and ensure that the unification takes place within the errors allowed by experimental measurements in the couplings. We find the unification scale and unified coupling to be $M_{\rm GUT}=6.3 \times 10^{16}$ GeV and $\alpha(M_{\rm GUT})=1/18$ respectively.

In nonsupersymmetric GUTs, the dominant contribution to proton decay arises through baryon and lepton number violating gauge interactions. Such gauge bosons typically have masses of the order of the GUT scale and hence the proton life time puts a lower bound on $M_{\rm GUT}$. The latest experimental limit on partial decay lifetime of proton is $\tau_p (p \to \pi^0 e^+) > 8.2 \times 10^{33}$ years \cite{Nishino:2009aa}. This implies
\be \label{pdecay}
M_{\rm GUT} \approx (m_p^5 ~\alpha(M_{\rm GUT})^2 ~\tau_p)^{1/4} \gsim 2.3 \times 10^{16} \sqrt{\alpha(M_{\rm GUT})}~{\rm GeV}~, 
\ee
where $m_p=0.938$ GeV is the proton mass. The values which we obtain for $M_{\rm GUT}$ and $\alpha(M_{\rm GUT})$ respect the above bounds. Further the predicted proton lifetime in this model is an order of magnitude higher than the current experimental limit and can be tested in the next generation proton decay experiments.

The color sextet fields which are introduced in order to achieve gauge coupling unification can naturally arise within SU(5) multiplets. For example $(\overline{6},1,-1/3)$ belongs to $45_H$ which is  already in the model to account for realistic fermion masses while $(\overline{6},3,-1/3)$ can come from a super heavy $50_H$ multiplet. In general, the masses of all the submultiplets are of the order of $M_{\rm GUT}$ and one needs to assume an additional fine tuning in parameters to keep some of the components light enough to achieve unification. Such an incomplete multiplet can also arise naturally without fine tuning in orbifold grand unified theories in which the gauge symmetry is broken through boundary condition in higher spacetime dimension, see for examples \cite{Kawamura:2000ev}. Note that the singlet scalar $S$ does not interact with these color sextet scalars belonging to $45_H$ and $50_H$ at tree level as it is forbidden by the SU(5) gauge symmetry.

\section{Summary and Outlook}
\label{summary}
The recent observation of 750 GeV diphoton excess by the ATLAS and CMS is an intriguing result. If it persists, it would be an insurmountable evidence of new physics beyond the SM. The observed cross section for this channel also provides a hint that 750 GeV resonance cannot be the only new physics particle. We need more colored/multiply-charged particles to explain the large production cross section/diphoton branching ratio respectively. We propose a simple nonsupersymmetric SU(5) grand unified model in which extra vector-like fermions naturally arise from a 24 dimensional adjoint representation of the gauge group. They possess appropriate color and electric charges leading to successful explanation for diphoton excess.  We find that the colored fermions $Q_8^0,~Q_3^{\pm 4/3}$ and $Q_3^{\pm 1/3}$ belonging to $24_F$ can enhance the production of singlet scalar $S$ while the electrically charged fermions $Q_3^{\pm 4/3},~Q_3^{\pm 1/3}$ and $L_3^\pm$ simultaneously enhance the $S\to \gamma\gamma$ branching ratio. We show that the observed diphoton cross section can be accounted in the model with sub-TeV adjoint fermions and with Yukawa coupling of ${\cal O}(1)$. For the adjoint fermion masses $M_f > 375$ GeV, a narrow width solution is preferred. A broad width scenario, as suggested by preliminary data, can be obtained if $Q_3^{\pm 1/3}$ is made lighter than 375 GeV and thereby opening the tree level decays of $S$. A simultaneous explanation of observed cross section and the total decay width in this case requires specific mass spectrum for adjoint fermions. The model is also shown to be consistent with the bounds on $S\to WW$, $S\to ZZ$ and $S\to Z\gamma$ decays.

The interactions of adjoint fermions with the SM fermions are forbidden using a discrete $Z_2$ symmetry under which $24_F$ is odd. Interestingly, this makes a singlet fermion residing in $24_F$ a candidate of cold dark matter. A successful explanation of diphoton anomaly in this model requires the mass of such dark matter particle in sub TeV range. The unification of gauge coupling is possible with light adjoint fermions and two colored sextet scalars leading to the values of unification scale and unified coupling while being consistent with the current proton lifetime limit.

\begin{acknowledgments} 
KMP thanks the Department of Science and Technology, Government of India for support under Inspire Faculty Award (INSPIRE-15-0088). 
\end{acknowledgments}


\begin{thebibliography}{99}

\bibitem{exp} 
LHC seminar \href{http://indico.cern.ch/event/442432/}{{\em ``ATLAS and CMS physics results from Run 2''}},
talks by Jim Olsen and Marumi Kado, CERN, 15 Dec. 2015.
ATLAS note,  ATLAS-CONF-2015-081, 
\href{https://atlas.web.cern.ch/Atlas/GROUPS/PHYSICS/CONFNOTES/ATLAS-CONF-2015-081/ATLAS-CONF-2015-081.pdf}{{\em ``Search for resonances decaying to photon pairs in 3.2 fb$^{-1}$ of $pp$
 collisions at $\sqrt{s}=13$ TeV with the ATLAS detector''}}.
CMS note, CMS PAS EXO-15-004 \href{https://cds.cern.ch/record/2114808/files/EXO-15-004-pas.pdf}{{\em ``Search for new physics in high mass diphoton events in proton-proton collisions at 13 TeV''}}.

\bibitem{Franceschini:2015kwy} 
  R.~Franceschini {\it et al.},
  JHEP {\bf 1603}, 144 (2016)
  [arXiv:1512.04933 [hep-ph]].
  
\bibitem{Gupta:2015zzs} 
  R.~S.~Gupta, S.~Jäger, Y.~Kats, G.~Perez and E.~Stamou,
  arXiv:1512.05332 [hep-ph].
  
\bibitem{singletscalars} 
  A.~Angelescu, A.~Djouadi and G.~Moreau,
  arXiv:1512.04921 [hep-ph];
  D.~Buttazzo, A.~Greljo and D.~Marzocca,
  arXiv:1512.04929 [hep-ph];
  S.~Di Chiara, L.~Marzola and M.~Raidal,
  arXiv:1512.04939 [hep-ph].
  
\bibitem{scalars} 
  B.~Bellazzini, R.~Franceschini, F.~Sala and J.~Serra,
  arXiv:1512.05330 [hep-ph];
  D.~Becirevic, E.~Bertuzzo, O.~Sumensari and R.~Z.~Funchal,
  arXiv:1512.05623 [hep-ph];
  S.~D.~McDermott, P.~Meade and H.~Ramani,
  arXiv:1512.05326 [hep-ph];
  M.~Low, A.~Tesi and L.~T.~Wang,
  arXiv:1512.05328 [hep-ph];
  A.~Kobakhidze, F.~Wang, L.~Wu, J.~M.~Yang and M.~Zhang,
  arXiv:1512.05585 [hep-ph];
  W.~Chao, R.~Huo and J.~H.~Yu,
  arXiv:1512.05738 [hep-ph];
  J.~Chakrabortty, A.~Choudhury, P.~Ghosh, S.~Mondal and T.~Srivastava,
  arXiv:1512.05767 [hep-ph];
  A.~Falkowski, O.~Slone and T.~Volansky,
  arXiv:1512.05777 [hep-ph];
  X.~F.~Han and L.~Wang,
  arXiv:1512.06587 [hep-ph];
  J.~Cao, C.~Han, L.~Shang, W.~Su, J.~M.~Yang and Y.~Zhang,
  arXiv:1512.06728 [hep-ph];
  I.~Chakraborty and A.~Kundu,
  arXiv:1512.06508 [hep-ph].

\bibitem{VLfermions} 
  J.~Ellis, S.~A.~R.~Ellis, J.~Quevillon, V.~Sanz and T.~You,
  arXiv:1512.05327 [hep-ph];
  B.~Dutta, Y.~Gao, T.~Ghosh, I.~Gogoladze and T.~Li,
  arXiv:1512.05439 [hep-ph];
  R.~Benbrik, C.~H.~Chen and T.~Nomura,
  arXiv:1512.06028 [hep-ph];
  W.~Liao and H.~q.~Zheng,
  arXiv:1512.06741 [hep-ph];
  G.~M.~Pelaggi, A.~Strumia and E.~Vigiani,
  arXiv:1512.07225 [hep-ph];
  C.~W.~Murphy,
  arXiv:1512.06976 [hep-ph];
  S.~M.~Boucenna, S.~Morisi and A.~Vicente,
  arXiv:1512.06878 [hep-ph].
    
\bibitem{others} 
  E.~Molinaro, F.~Sannino and N.~Vignaroli,
  arXiv:1512.05334 [hep-ph];
  A.~Pilaftsis,
  arXiv:1512.04931 [hep-ph];
  S.~Knapen, T.~Melia, M.~Papucci and K.~Zurek,
  arXiv:1512.04928 [hep-ph];
  M.~Backovic, A.~Mariotti and D.~Redigolo,
  arXiv:1512.04917 [hep-ph];
  X.~J.~Bi, Q.~F.~Xiang, P.~F.~Yin and Z.~H.~Yu,
  arXiv:1512.06787 [hep-ph];
  Y.~Mambrini, G.~Arcadi and A.~Djouadi,
  arXiv:1512.04913 [hep-ph];
  K.~Harigaya and Y.~Nomura,
  arXiv:1512.04850 [hep-ph];
  Y.~Nakai, R.~Sato and K.~Tobioka,
  arXiv:1512.04924 [hep-ph];
  T.~Higaki, K.~S.~Jeong, N.~Kitajima and F.~Takahashi,
  arXiv:1512.05295 [hep-ph];
  C.~Petersson and R.~Torre,
  arXiv:1512.05333 [hep-ph];
  P.~Cox, A.~D.~Medina, T.~S.~Ray and A.~Spray,
  arXiv:1512.05618 [hep-ph];
  A.~Ahmed, B.~M.~Dillon, B.~Grzadkowski, J.~F.~Gunion and Y.~Jiang,
  arXiv:1512.05771 [hep-ph];
  J.~M.~No, V.~Sanz and J.~Setford,
  arXiv:1512.05700 [hep-ph];
  S.~Fichet, G.~von Gersdorff and C.~Royon,
  arXiv:1512.05751 [hep-ph];
  C.~Csaki, J.~Hubisz and J.~Terning,
  arXiv:1512.05776 [hep-ph];
  E.~Gabrielli, K.~Kannike, B.~Mele, M.~Raidal, C.~Spethmann and H.~Veermäe,
  arXiv:1512.05961 [hep-ph];
  E.~Megias, O.~Pujolas and M.~Quiros,
  arXiv:1512.06106 [hep-ph];
  L.~M.~Carpenter, R.~Colburn and J.~Goodman,
  arXiv:1512.06107 [hep-ph];
  C.~Han, H.~M.~Lee, M.~Park and V.~Sanz,
  arXiv:1512.06376 [hep-ph];
  S.~Chang,
  arXiv:1512.06426 [hep-ph];
  H.~Han, S.~Wang and S.~Zheng,
  arXiv:1512.06562 [hep-ph];
  D.~Bardhan, D.~Bhatia, A.~Chakraborty, U.~Maitra, S.~Raychaudhuri and T.~Samui,
  arXiv:1512.06674 [hep-ph];
  T.~F.~Feng, X.~Q.~Li, H.~B.~Zhang and S.~M.~Zhao,
  arXiv:1512.06696 [hep-ph];
  D.~Barducci, A.~Goudelis, S.~Kulkarni and D.~Sengupta,
  arXiv:1512.06842 [hep-ph];
  R.~Ding, L.~Huang, T.~Li and B.~Zhu,
  arXiv:1512.06560 [hep-ph];
  O.~Antipin, M.~Mojaza and F.~Sannino,
  arXiv:1512.06708 [hep-ph];
  J.~M.~Cline and Z.~Liu,
  arXiv:1512.06827 [hep-ph];
  M.~Bauer and M.~Neubert,
  arXiv:1512.06828 [hep-ph];
  M.~T.~Arun and P.~Saha,
  arXiv:1512.06335 [hep-ph];
  P.~S.~B.~Dev and D.~Teresi,
  arXiv:1512.07243 [hep-ph];
  U.~K.~Dey, S.~Mohanty and G.~Tomar,
  arXiv:1512.07212 [hep-ph];
  A.~Alves, A.~G.~Dias and K.~Sinha,
  arXiv:1512.06091 [hep-ph];
  W.~S.~Cho, D.~Kim, K.~Kong, S.~H.~Lim, K.~T.~Matchev, J.~C.~Park and M.~Park,
  arXiv:1512.06824 [hep-ph];
  J.~J.~Heckman,
  arXiv:1512.06773 [hep-ph];
  M.~Chala, M.~Duerr, F.~Kahlhoefer and K.~Schmidt-Hoberg,
  arXiv:1512.06833 [hep-ph];
  J.~de Blas, J.~Santiago and R.~Vega-Morales,
  arXiv:1512.07229 [hep-ph];
  Q.~H.~Cao, Y.~Liu, K.~P.~Xie, B.~Yan and D.~M.~Zhang,
  arXiv:1512.05542 [hep-ph];
  W.~Chao,
  arXiv:1512.06297 [hep-ph];
  F.~P.~Huang, C.~S.~Li, Z.~L.~Liu and Y.~Wang,
  arXiv:1512.06732 [hep-ph];
  L.~Bian, N.~Chen, D.~Liu and J.~Shu,
  arXiv:1512.05759 [hep-ph].
  
\bibitem{Georgi:1974sy} 
  H.~Georgi and S.~L.~Glashow,
  Phys.\ Rev.\ Lett.\  {\bf 32}, 438 (1974).



\bibitem{Georgi:1979df} 
  H.~Georgi and C.~Jarlskog,
  Phys.\ Lett.\ B {\bf 86}, 297 (1979).

  
  
\bibitem{Perez:2007rm} 
  P.~Fileviez Perez,
  Phys.\ Lett.\ B {\bf 654}, 189 (2007)
  [hep-ph/0702287];
  I.~Dorsner and P.~Fileviez Perez,
  Phys.\ Lett.\ B {\bf 642}, 248 (2006)
  [hep-ph/0606062].
  
  
\bibitem{Dorsner:2005fq} 
  I.~Dorsner and P.~Fileviez Perez,
  Nucl.\ Phys.\ B {\bf 723}, 53 (2005)
  [hep-ph/0504276];
  I.~Dorsner, P.~Fileviez Perez and R.~Gonzalez Felipe,
  Nucl.\ Phys.\ B {\bf 747}, 312 (2006)
  [hep-ph/0512068].
  
\bibitem{bound:Shh}  
  G.~Aad {\it et al.} [ATLAS Collaboration],
  Phys.\ Rev.\ D {\bf 92}, 092004 (2015)
  [arXiv:1509.04670 [hep-ex]];
  V.~Khachatryan {\it et al.} [CMS Collaboration],
  Phys.\ Lett.\ B {\bf 749}, 560 (2015)
  [arXiv:1503.04114 [hep-ex]].
  
\bibitem{Bajc:2006ia} 
  B.~Bajc and G.~Senjanovic,
  JHEP {\bf 0708}, 014 (2007)
  [hep-ph/0612029].

    
\bibitem{Djouadi:2005gi} 
  A.~Djouadi,
  Phys.\ Rept.\  {\bf 457}, 1 (2008)
  [hep-ph/0503172];
  Phys.\ Rept.\  {\bf 459}, 1 (2008)
  [hep-ph/0503173].

\bibitem{mg5} 
  J.~Alwall {\it et al.},
  JHEP {\bf 1407}, 079 (2014)
  [arXiv:1405.0301 [hep-ph]].

  
\bibitem{feynrules} 
  A.~Alloul, N.~D.~Christensen, C.~Degrande, C.~Duhr and B.~Fuks,
  Comput.\ Phys.\ Commun.\  {\bf 185}, 2250 (2014)
  [arXiv:1310.1921 [hep-ph]].

\bibitem{Ellis:1975ap} 
  J.~R.~Ellis, M.~K.~Gaillard and D.~V.~Nanopoulos,
  Nucl.\ Phys.\ B {\bf 106}, 292 (1976);
  M.~Carena, I.~Low and C.~E.~M.~Wagner,
  JHEP {\bf 1208}, 060 (2012)
  [arXiv:1206.1082 [hep-ph]].

  
  
\bibitem{Chatrchyan:2013oca} 
  S.~Chatrchyan {\it et al.} [CMS Collaboration],
  JHEP {\bf 1307}, 122 (2013)
  [arXiv:1305.0491 [hep-ex]].
  
\bibitem{Langacker:1991an} 
  P.~Langacker and M.~x.~Luo,
  Phys.\ Rev.\ D {\bf 44}, 817 (1991).
  
\bibitem{Patel:2011eh} 
  K.~M.~Patel and P.~Sharma,
  JHEP {\bf 1104}, 085 (2011)
  [arXiv:1102.4736 [hep-ph]].
  
\bibitem{Nishino:2009aa} 
  H.~Nishino {\it et al.} [Super-Kamiokande Collaboration],
  Phys.\ Rev.\ Lett.\  {\bf 102}, 141801 (2009)
  [arXiv:0903.0676 [hep-ex]].
  
\bibitem{Kawamura:2000ev} 
  Y.~Kawamura,
  Prog.\ Theor.\ Phys.\  {\bf 105}, 999 (2001)
  [hep-ph/0012125];
  G.~Altarelli and F.~Feruglio,
  Phys.\ Lett.\ B {\bf 511}, 257 (2001)
  [hep-ph/0102301];
  F.~Feruglio, K.~M.~Patel and D.~Vicino,
  JHEP {\bf 1509}, 040 (2015)
  [arXiv:1507.00669 [hep-ph]].
  
\end{thebibliography}
\end{document}